\documentclass[smallextended]{svjour3}       
\smartqed  

\usepackage[english]{babel}
\usepackage{amsmath}
\usepackage{amssymb}
\usepackage{dsfont}
\usepackage{float} 
\usepackage{enumitem}
\usepackage{multirow}

\input{qcircuit} 
\def\<{\langle}\def\>{\rangle}
\def\n#1{|\!|#1|\!|}

\def\Uset{\mathbb{U}}	
\def\Supp{\mathsf{Supp}\,}

\def\tA{\transf{A}}
\def\tU{\transf{U}}\def\tT{\transf{T}}

\def\rA{\sys{A}}\def\rB{\sys{B}}\def\rC{\sys{C}}
\def\rD{\sys{D}}\def\rI{\sys{I}}
\def\rE{\sys{E}}\def\rF{\sys{F}}

\def\<{\langle}\def\>{\rangle}\def\kk{\>\! \>}\def\bb{\<\!\<}

\def\trnsfrm#1{\mathcal #1}
\def\tA{\trnsfrm A} 
 
\def\tT{\trnsfrm T}\def\tU{\trnsfrm U}

\def\tP{\trnsfrm P } \def\tT{\trnsfrm T} 
\def\rA{{\rm A}}\def\rB{{\rm B}}\def\rC{{\rm C}}\def\rD{{\rm D}} \def\rE{{\rm E}} \def\rF{{\rm F}}
\def\rI{{\rm I}}

\def\PurSt{\rm{PurSt}}\def\St{\rm{St}}\def\Eff{\mathrm{Eff}}
\def\Trn{\mathrm{Transf}}

\def\Tr{\operatorname{Tr}}
\def\Hyp{{\rm Hyp}}\def\TRUE{\mathsf {TRUE}}\def\FALSE{\mathsf {FALSE}}
\def\Set{{\mathsf S}}
\def\Conv{\mathsf{Conv}}
\def\Cone{\mathsf{Cone}}

\def\trnsfrm#1{\mathcal #1}
\def\tA{\trnsfrm A} 
 
\def\tT{\trnsfrm T}\def\tU{\trnsfrm U}

\def\tP{\trnsfrm P } \def\tT{\trnsfrm T} 
\def\rA{{\rm A}}\def\rB{{\rm B}}\def\rC{{\rm C}}\def\rD{{\rm D}} \def\rE{{\rm E}} \def\rF{{\rm F}}
\def\rI{{\rm I}}

\def\sH{{\mathcal{H}}}\def\sK{{\mathcal{K}}}\def\sS{{\mathcal{S}}}
\def\St{\rm{St}}\def\Eff{\mathrm{Eff}}\def\Trn{\mathrm{Trn}}\def\T{\mathrm{T}}\def\Bnd{\mathrm{Bnd}}\def\Bnd{\mathrm{B}}
\def\Cmplx{\mathbb{C}}\def\Reals{\mathbb{R}}
\def\CP{\mathrm{CP}}\def\P{\mathrm{P}}
\def\HS{\mathrm{HS}}
\def\BSt{\rm{\partial\St}}

\def\transp#1{{#1}^\intercal}
\def\eg{{e.~g.} }
\def\rnk{\operatorname{rank}}
\def\Mrkv{\mathrm{Mrkv}}\def\Prm{\mathrm{Prm}}
\def\vx{{\bf x}}

\journalname{FOUNDATIONS OF PHYSICS}
\begin{document}

\title{QUANTUM EPISTEMOLOGY AND FALSIFICATION OF RANDOM GENERATORS\thanks{This work was made possible through the support of the Elvia and Federico Faggin Foundation, Grant 2020-214365.}}

\titlerunning{QUANTUM EPISTEMOLOGY}

\author{Giacomo Mauro D'Ariano}
\institute{Dipartimento di Fisica\\ dell'Universit\`a di Pavia \at
              via Bassi 6, 27100 Pavia\\
              Tel.: +39 347 0329998\\
              \email{dariano@unipv.it}\\
             \emph{Also:}  Istituto Lombardo Accademia di Scienze e Lettere\\
		INFN, Gruppo IV, Sezione di Pavia
}

\date{Received: date / Accepted: date}

\maketitle

\begin{abstract}
The operational axiomatization of quantum theory in \cite{purification,QUIT-Arxiv,CDPbook} can be regarded as a set of six epistemological rules for falsifying propositions of the theory. In particular, the Purification postulate--the only one that is not shared with classical theory--allows falsification of random-sequences generators, a task unfeasible classically.\end{abstract}

\section{Introduction}

Our physical world is ruled by two theories: classical theory (CT) and quantum theory (QT).
Compared to CT, QT still looks weird: however, this may be  the symptom that we are still missing the hidden logic of the theory. Indeed, we should not forget that among the two  theories QT is the most powerful one, simply because CT is a restriction of QT. In fact, for given system dimension $d$, CT restricts QT's states to the convex hull of a fixed maximal set of  jointly perfectly discriminable pure states (the $d$-simplex), and, correspondingly, transformations are restricted to (sub)Markov linear maps.\footnote{For mathematical axiomatizations and main theorems of both theories QT and CT see Appendice \ref{appendices}.}  We can thus regard the indeterminism inherent QT as the price to be payed for adding information-processing power.

\medskip
Deriving QT from information-theoretical principles  \cite{purification,QUIT-Arxiv,CDPbook} reveals how the theory is more powerful than CT. Indeed, the two theories share five postulates, whereas the sixth QT postulate highlights the fundamental task that QT can achieve whereas CT cannot: {\em purification}. On the other hand, the sixth CT postulate makes the theory a restriction of QT .  Thus purification synthesizes the additional power of  QT compared to CT.

\medskip 
In the convex-OPT\footnote{OPT is the acronym for Operational Probabilistic Theory. See also Table \ref{tnotat} in Appendix \ref{appendices} for acronyms, abbreviations, and symbols. For OPTs see any of Refs.\cite{purification,QUIT-Arxiv,CDPbook}.} language, the five common postulates are:

\begin{enumerate}[label=P\arabic*]
\item {\em Causality:} the probability of preparation is independent on the choice of observation. 
\item {\em Perfect discriminability:} every state on the boundary of the convex set of states can be perfectly distinguished from some other state.
\item {\em Local discriminability:} It is possible to discriminate any pair of states of composite systems using only local observations.
\item {\em Compressibility:} For all states which are not completely mixed there exists an ideal compression scheme.
\item {\em Atomicity of composition:} The composition of two atomic transformations is atomic.
\end{enumerate}

The sixth postulates, different for the two theories, are:
\begin{enumerate}[label=P\arabic*]
\item[P6$_Q$] {\em Purification:} Every state has a purification. For fixed purifying system, every two purifications of the same state are connected by a reversible transformation on the purifying system. 
\item[P6$_C$] {\em Perfect joint discrimination:} For any system all pure states can be perfectly discriminated jointly.
\end{enumerate}
Notice that P6$_C$ forces CT to restrict QT's pure states to a maximal set of perfectly discriminable ones.

\section{The Purification Principle}\label{sec:purprinc}
Let's recall the statement of the principle.
\par\vskip 10pt {\em For every system $\rA$ and for every state $\rho\in \St(\rA)$, there exists a system $\rB$ and a pure state $ \Psi \in \PurSt (\rA \rB)$ such that
 \begin{equation}\label{existence}
\begin{aligned} \Qcircuit @C=1em @R=.7em @! R {\prepareC{\rho}& \qw \poloFantasmaCn \rA &\qw } \end{aligned}
~=~
\begin{aligned}
\Qcircuit @C=1em @R=.7em @! R {\multiprepareC{1}{\Psi}& \qw \poloFantasmaCn \rA &\qw \\
\pureghost{\Psi} & \qw \poloFantasmaCn \rB &\measureD {e}}  
\end{aligned}  \quad  \, .
\end{equation} 
If two pure states $\Psi$ and $\Psi'$ satisfy 
\begin{equation*}
\begin{aligned}
\Qcircuit @C=1em @R=.7em @! R {\multiprepareC{1}{\Psi'}& \qw \poloFantasmaCn \rA &\qw \\
\pureghost{\Psi'} & \qw \poloFantasmaCn \rB &\measureD {e}}  
\end{aligned}  ~=~  
\begin{aligned}
\Qcircuit @C=1em @R=.7em @! R {\multiprepareC{1}{\Psi}& \qw \poloFantasmaCn \rA &\qw \\
\pureghost{\Psi} & \qw \poloFantasmaCn \rB &\measureD {e}}  
\end{aligned}   \quad ,
\end{equation*} 
then there exists a reversible transformation $\tU$, acting only on system $\rB$, such that 
\begin{equation}\label{uniqueness}
\begin{aligned}
\Qcircuit @C=1em @R=.7em @! R {\multiprepareC{1}{\Psi'}& \qw \poloFantasmaCn \rA &\qw \\
\pureghost{\Psi'} & \qw \poloFantasmaCn \rB &\qw}  
\end{aligned}  ~=~  
\begin{aligned}
\Qcircuit @C=1em @R=.7em @! R {\multiprepareC{1}{\Psi}& \qw \poloFantasmaCn \rA &\qw &\qw&\qw \\
\pureghost{\Psi} & \qw \poloFantasmaCn \rB &\gate{\tU}   &\qw \poloFantasmaCn \rB  &\qw}  
\end{aligned}   \quad .
\end{equation} }

We call $\Psi$ \emph{a purification of} $\rho$, with $\rB$ \emph{purifying system}.  

Informally, Eq. (\ref{existence}) guarantees that we can always find a pure state of $\rA\rB$ that is compatible with our limited knowledge of $\rA$ alone. Furthermore, Eq. (\ref{uniqueness})  specifies that all the states of $\rA\rB$ that are compatible with our knowledge of $\rA$ are essentially the same, up to a reversible transformation on $\rB$.    We will  call this property  \emph{uniqueness of purification}.\footnote{Note that the two purifications in Eq. (\ref{uniqueness})   have the same purifying system.}  

\section{Epistemological value of the postulates\label{Erul}}
In quantum logic \cite{QL} one associates a "proposition" about the system $\rA$ to an orthogonal projector $P_\sS$ on a subspace $\sS\subseteq\sH_\rA$ of  the Hilbert space of $\rA$. Being the map between $P_\sS$ and its support $\sS$ \footnote{Here, by "support" of an operator we mean the orthogonal complement of its kernel.} a bijection, one can equivalently  associate "propositions" to Hilbert subspaces $\sS\subseteq\sH_\rA$. One can now enrich the notion of "proposition" by associating it to a quantum state $\rho$ with support $\Supp\rho=\sS$, the state $\rho$ encoding a reacher information than just its support $\sS$. We conclude that the notion of "state"  constitutes a more detailed concept of "proposition" than that of orthogonal projector. For such a reason in the present context it is more appropriate to associate the word "proposition" to the notion of "quantum state" instead of the original definition as orthogonal projector. In conclusion "proposition about $\rA$" will be synonym of "state of $\rA$". Clearly, the same notion can be extended to CT, since CT is a restriction of QT. 

According to the above identification we can now translate QT and CT into the language of "propositions", and appreciate how, remarkably, all the six postulates for QT and CT are all of epistemological nature: they all are assertions regarding falsifiability of theory's "propositions".

\medskip
\subsection*{Epistemological rules: }
\begin{enumerate}[label=E\arabic*]
\item Causality is required for falsification of propositions by observations.
\item Perfect discriminability guarantees the existence of falsifiable propositions\footnote{Notice the abundance of falsifiable propositions, since they make the boundary of the convex set of states (internal states are completely mixed).} derived from the theory. 
\item Local discriminability guarantees that falsifiability of joint propositions can be accomplished locally.\footnote{Here locally means using single system observations and classical communication.}
\item Compressibility provides the possibility of reducing dimension of the system supporting a falsifiable proposition.  
\item Atomicity of composition guarantees the existence of a class of transformations that do not affect falsifiability of propositions. 
\item[E6$_Q$] Purification allows for falsifiable random generators. 
\item[E6$_C$] Perfect discrimination allows any set of determinate propositions to be jointly falsifiable.
\end{enumerate}

\smallskip
Any of the above statements constitutes an epistemological power of the corresponding principle. Statement E6$_Q$ in particular establishes the possibility of  logically falsifying a quantum random generator, namely, there exists a falsification test that establishes if a given quantum random generator is different from a claimed one. Notice that such a  falsification cannot be achieved classically for probabilities that are not deterministic, i.e. $p\notin\{0,1\}$:\footnote{Classically one falsifies the generator for $p=1$ whenever the event does not occur, and  for $p=0$ when it does. In such a case the probability value itself is directly falsified. We stress that in the other cases QT makes falsifiable the random generator--not the probability value.} no succession of outcomes can logically falsify a value of $p$ different from 0 and 1. Remarkably, as we will see in this paper, thanks to the purification postulate, within QT we can falsify random generators with any probability distribution. 

Before proving the epistemological rules, we recall the theory of falsification tests introduced in Ref. \cite{DAriano-nopuri}

\section{The falsification test}
We say that an event $F$ is a {\em falsifier} of hypothesis $\Hyp$ if $F$ cannot happen for $\Hyp=\TRUE$. We will call the binary test $\{F,F_?\}$ {\em a falsification test} for hypothesis $\Hyp$, and denote by $F_?$ the {\em inconclusive event}. Notice that the occurrence of $F_?$ generally does not mean that $\Hyp=\TRUE$, but that {\Hyp}  has not been falsified.
\par
Suppose now that one wants to falsify a proposition about the quantum state $\rho\in\St(\rA)$ of system $\rA$. In such case any effective falsification test  can be achieved as a binary {\em observation test} of the form
\begin{equation}
\{F,F_?\}\subset\Eff(A),\quad F_?:=I_\rA-F,\quad F>0,F_?\geq 0,
\end{equation}
where by the symbol $F$ ($F_?$) we denote both event and corresponding positive operator. The strict positivity of $F$ is required for effectiveness of the test,  $F=0$ corresponding to the {\em inconclusive test}, which outputs only the inconclusive outcome. On the other hand, $F_?=0$ corresponds to the logical {\em a priori} falsification. 

Examples of inconclusive tests have been given in Ref. \cite{DAriano-nopuri}, to prove that hypotheses as "purity of an unknown state", or "unitarity of an unknown transformation" cannot be falsified.
\subsection{Falsification of a quantum state support}
Consider  the proposition
\begin{equation}\quad\label{assert}
\Hyp:\quad\Supp\rho=:\sK\subsetneq\sH_\rA,\;\rho\in\St(\rA),\qquad\dim\sH_\rA\geq 2
\end{equation}
$\Supp\rho$ denoting the support of  $\rho$. Then,  any operator of the form
\begin{equation}\label{falsifier}
0<F\leqslant I_\rA,\quad \Supp F\subseteq\sK^\perp
\end{equation}
would have zero expectation for a state $\rho$ satisfying $\Hyp $ in Eq. \eqref{assert}, which means that occurrence  of $F$ would falsify   \Hyp, namely
\begin{equation}
\Tr[\rho F]>0\,\Rightarrow \Hyp=\FALSE.
\end{equation}
Eq. \eqref{falsifier} provides the most general falsification test of  $\Hyp$ in Eq. \eqref{assert}, the choice $\Supp F=\sK^\perp$ corresponding to the most efficient test, namely the one maximizing  falsification chance.  Notice that the outcome corresponding to $I-F$ does not correspond to a verification of  \Hyp, since it generally  can occur for  $\Supp(I-F)\cap\sK\neq\emptyset$.

\section{Proofs of  epistemological rules}
In this section we prove the epistemological rules given in Sect. \ref{Erul}. We will denote by $\St(\rA)$ the convex set of states of system $\rA$  and by $\BSt(\rA)$ its boundary.

\begin{enumerate}[label=E\arabic*]
\item For any proposition $\rho$ that is falsifiable (i.e. $\rnk\rho<d_\rA$), causality protects the falsification target state to be changed by the particular choice of observation.
\item Consider two quantum states $\rho,\nu\in\St(\rA)$. They are perfectly discriminable iff  $\Supp\rho\perp\Supp\nu$, which implies that $2\leq\rnk\rho+\rnk\nu\leq d_\rA$ with both ranks at least unit. It follows that $1\leq\rnk\rho\leq d_\rA-\min\rnk\nu=d_\rA-1$ and the same for $\rnk\nu$, hence both $\rho,\nu\in\BSt(\rA)$. We conclude that the two states are falsifiable, with falsifiers $\ker\rho\subseteq\Supp\nu$ and $\ker\nu\subseteq\Supp\rho$, respectively.\smallskip
\item Consider the pure entangled state corresponding to state-vector 
$|A\kk\in\sH_\rA\otimes\sH_B $
\footnote{Here we are using the {\em double-ket} notation \cite{bellobs} (for a thorough treatment see \cite{CDPbook}). Once it is chosen the orthonormal factorized canonical basis $\{|i\>\otimes|j\>\}$ for $\sH\otimes\sH$, we have the one-to-one  correspondence between vectors in $\sH\otimes\sH$ and operators on $\sH$ 
\begin{equation}\nonumber
|\Psi\kk:=\sum_{ij}\Psi_{ij}|i\>\otimes|j\>\;\longleftrightarrow\quad\Psi=\sum_{ij}\Psi_{ij}|i\>\<j|\in \HS(\Cmplx^d),
\end{equation}
where $\HS(\Cmplx^d)$ denotes Hilbert-Schmidt operators in dimensions $d$. One then can veryify the following identity
\begin{equation}\nonumber
(A\otimes B)|C\kk=|AC\transp{B}\kk,
\end{equation}
$\transp{B}$ denoting the {\em transposed operator} of $B$, the operator that has the transposed matrix w.r.t. the canonical basis. Notice that \eg $\transp{(|\phi\>\<\psi|)}=|\psi^*\>\<\phi^*|$ where $|\psi^*\>$ is the vector $|\psi\>$ with complex-conjugated coefficients w.r.t. the canonical basis.}
The following sequence of identities holds
\begin{equation}\label{liked}
(\<a|\otimes\<b|)|A\kk=(\<a|\otimes I)(I\otimes\<b|A^T)\sum_n|a_n\>\otimes| a_n\>^*=\<b|A^T(|a\>^*)=\<b|(A^\dag|a\>)^*
\end{equation}
where $\Set:=\{|a_n\>\}_{n=1}^{d_\rA}$ is an orthonormal basis for $\sH_\rA$, with $|a\>\in\Set$.  Eq. \ref{liked} shows that choosing $|b\>$ orthogonal to  $(A^\dag|a\>)^*$ one has local falsifier of the entangled state $|A\kk\bb A|$ given by 
\begin{equation}
P_\rA\otimes P_\rB=|a\>\<a|\otimes|b\>\<b|,
\end{equation}
The generalization to general mixtures $R$ is straightforward, upon writing the state $R$ in the canonical form 
\begin{equation}
R=\sum_{j=1}^{d_\rA}|A_j\kk\bb A_j|,\quad \Tr(A_i^\dag A_j)=
\delta_{ij}p_j,\;\sum_{j=1}^{d_\rA} p_j=1.
\end{equation} 

\medskip
\item Any falsifiable state $\rho$ has $\dim\ker\rho\geq1$, hence it can be isometrically mapped to a state of a system $\rB$ with $d_\rB\leq d_\rA-1$.
\item A transformation $\tA\in\Trn(\rA\!\!\to\!\!\rB)$ is called {\em atomic} if it has only one Krauss term, namely it can be written as $\tA\rho=A\rho A^\dag$, with $A\in\Bnd(\sH_\rA)$ and $\n{A}\leq 1$. This implies that $\rnk(\tA\rho)\leq\rnk\rho$, namely the falsification space has dimension which is not decreased, hence the output state $\tA\rho$ can be falsified. This is not necessarily true for  $\tA$ {\em non atomic}, namely with more than one Krauss term, i.~e. $\tA\rho=A_1\rho A_1^\dag  +A_2\rho A_2^\dag+\ldots$.
\item[E6$_Q$] See Sect.\ref{FBQC}
\item[E6$_C$] It trivially holds for CT.
\end{enumerate}

\section{Falsification of a biased quantum coin.}\label{FBQC}
A falsifiable setup for a quantum binary random generator can use any quantum system $\rA$,  e.~g.  a qubit, in a pure state $\rho=|\psi\>\<\psi|$ along with  an orthogonal observation test $\Omega=\{\omega_i\}_{i=0,1}$ with $\omega_i=|i\>\<i|$, $\< i|j\>=\delta_{ij}$ (namely a customarydiscrete observable). The following setup
\begin{equation}
i\in\{0,1\},\quad p_i=\Qcircuit @C=1em @R=.7em @! R {
    \prepareC{\rho}&\ustick{\rA}\qw&\measureD{\omega _i}},\quad
    \rho=|\psi\>\<\psi|,\;|\psi\>=\sqrt{p}|0\>+\sqrt{1-p}e^{i\phi}|1\>,\;\omega_i=|i\>\<i|, \quad\;
    \end{equation}
 is a binary random generator with probabilities with $p_0=p$.

The advantage of this choice of setup (compared e.g. to using a mixed state and/or a non orthogonal observation test) is that it can be falsifiable. In fact, the preparation of the state $|\psi\>\<\psi|$ can be falsified efficiently by running the falsification test of the state support, using falsifier $F=I-|\psi\>\<\psi|$. On the other hand, the observation test $\Omega$ can be taken as just the observable providing the physical meaning of the orthonormal basis chosen for the qubit $\rA$ (e. g. spin-up and spin-down), which is required to define physically the state preparation.
\footnote{The above setup can be trivially generalized to a $N$-ary random generator ($N>2$) with probability distribution $\{p_n\}_{n\in\mathds{Z}_N}$ by using a system $\rA$ with $\dim(\sH_\rA)=N$, and a pure state with vector with more than two nonvanishing probability amplitudes. Here, $F=I-|\psi\>\<\psi|$ still provides the most efficient falsifier. Notice that for $\dim(\sH_\rA)>2$ it is also possible to falsify mixed states with rank strictly smaller than $\dim(\sH_\rA)$.}
\footnote{Notice that the probability of falsification of a mixed state $\rho\neq |\psi\>\<\psi|$  is given by $p=1-\<\psi|\rho|\psi\>$, and vanishes linearly with the overlap between the declared state $|\psi\>\<\psi|$ and the true state $\rho$.}

\section{Conclusions}
CT and QT are more than theories about the world: as OPTs they constitute extensions of logic.\footnote{Famously von Neumann attempted to prove QT be a kind of logic. We now know that, instead, it is an extension of it.} We have seen that QT can be regarded as a chapter of epistemology, being a set of rules for accessibility of falsifications. Thus QT, more than answering the question “what is reality”, it provides rules for “how we can explore it”.  
One then can add axioms to those of QT to get more refined theories, such as Free Quantum Field Theory. The latter can be indeed obtained upon considering a denumerable set of QT systems, and adding the axioms of locality, homogeneity, and isotropy of interactions (see e.g. the review \cite{PWP}).
The advantage of this choice of setup (compared e.g. to using a mixed state and/or a non orthogonal observation test) is that it can be falsifiable. In fact, the preparation of the state $|\psi\>\<\psi|$ can be falsified efficiently by running the falsification test of the state support, using falsifier $F=I-|\psi\>\<\psi|$. On the other hand, the observation test $\Omega$ can be taken as just the observable providing the physical meaning of the orthonormal basis chosen for the qubit $\rA$ (e. g. spin-up and spin-down), which is required to define physically the state preparation.
\footnote{The above setup can be trivially generalized to a $N$-ary random generator ($N>2$) with probability distribution $\{p_n\}_{n\in\mathds{Z}_N}$ by using a system $\rA$ with $\dim(\sH_\rA)=N$, and a pure state with vector with more than two nonvanishing probability amplitudes. Here, $F=I-|\psi\>\<\psi|$ still provides the most efficient falsifier. Notice that for $\dim(\sH_\rA)>2$ it is also possible to falsify mixed states with rank strictly smaller than $\dim(\sH_\rA)$.}
\footnote{Notice that the probability of falsification of a mixed state $\rho\neq |\psi\>\<\psi|$  is given by $p=1-\<\psi|\rho|\psi\>$, and vanishes linearly with the overlap between the declared state $|\psi\>\<\psi|$ and the true state $\rho$.}

\section{Conclusions}
CT and QT are more than theories about the world: as OPTs they constitute extensions of logic.\footnote{Famously von Neumann attempted to prove QT be a kind of logic. We now know that, instead, it is an extension of it.} We have seen that QT can be regarded as a chapter of epistemology, being a set of rules for accessibility of falsifications. Thus QT, more than answering the question “what is reality”, it provides rules for “how we can explore it”.  
One then can add axioms to those of QT to get more refined theories, such as Free Quantum Field Theory. The latter can be indeed obtained upon considering a denumerable set of QT systems, and adding the axioms of locality, homogeneity, and isotropy of interactions (see e.g. the review \cite{PWP}).
The advantage of this choice of setup (compared e.g. to using a mixed state and/or a non orthogonal observation test) is that it can be falsifiable. In fact, the preparation of the state $|\psi\>\<\psi|$ can be falsified efficiently by running the falsification test of the state support, using falsifier $F=I-|\psi\>\<\psi|$. On the other hand, the observation test $\Omega$ can be taken as just the observable providing the physical meaning of the orthonormal basis chosen for the qubit $\rA$ (e. g. spin-up and spin-down), which is required to define physically the state preparation.
\footnote{The above setup can be trivially generalized to a $N$-ary random generator ($N>2$) with probability distribution $\{p_n\}_{n\in\mathds{Z}_N}$ by using a system $\rA$ with $\dim(\sH_\rA)=N$, and a pure state with vector with more than two nonvanishing probability amplitudes. Here, $F=I-|\psi\>\<\psi|$ still provides the most efficient falsifier. Notice that for $\dim(\sH_\rA)>2$ it is also possible to falsify mixed states with rank strictly smaller than $\dim(\sH_\rA)$.}
\footnote{Notice that the probability of falsification of a mixed state $\rho\neq |\psi\>\<\psi|$  is given by $p=1-\<\psi|\rho|\psi\>$, and vanishes linearly with the overlap between the declared state $|\psi\>\<\psi|$ and the true state $\rho$.}

\section{Conclusions}
CT and QT are more than theories about the world: as OPTs they constitute extensions of logic.\footnote{Famously von Neumann attempted to prove QT be a kind of logic. We now know that, instead, it is an extension of it.} We have seen that QT can be regarded as a chapter of epistemology, being a set of rules for accessibility of falsifications. Thus QT, more than answering the question “what is reality”, it provides rules for “how we can explore it”.  
One then can add axioms to those of QT to get more refined theories, such as Free Quantum Field Theory. The latter can be indeed obtained upon considering a denumerable set of QT systems, and adding the axioms of locality, homogeneity, and isotropy of interactions (see e.g. the review \cite{PWP}).

The advantage of this choice of setup (compared e.g. to using a mixed state and/or a non orthogonal observation test) is that it can be falsifiable. In fact, the preparation of the state $|\psi\>\<\psi|$ can be falsified efficiently by running the falsification test of the state support, using falsifier $F=I-|\psi\>\<\psi|$. On the other hand, the observation test $\Omega$ can be taken as just the observable providing the physical meaning of the orthonormal basis chosen for the qubit $\rA$ (e. g. spin-up and spin-down), which is required to define physically the state preparation.
\footnote{The above setup can be trivially generalized to a $N$-ary random generator ($N>2$) with probability distribution $\{p_n\}_{n\in\mathds{Z}_N}$ by using a system $\rA$ with $\dim(\sH_\rA)=N$, and a pure state with vector with more than two nonvanishing probability amplitudes. Here, $F=I-|\psi\>\<\psi|$ still provides the most efficient falsifier. Notice that for $\dim(\sH_\rA)>2$ it is also possible to falsify mixed states with rank strictly smaller than $\dim(\sH_\rA)$.}
\footnote{Notice that the probability of falsification of a mixed state $\rho\neq |\psi\>\<\psi|$  is given by $p=1-\<\psi|\rho|\psi\>$, and vanishes linearly with the overlap between the declared state $|\psi\>\<\psi|$ and the true state $\rho$.}

\section{Conclusions}
CT and QT are more than theories about the world: as OPTs they constitute extensions of logic.\footnote{Famously von Neumann attempted to prove QT be a kind of logic. We now know that, instead, it is an extension of it.} We have seen that QT can be regarded as a chapter of epistemology, being a set of rules for accessibility of falsifications. Thus QT, more than answering the question “what is reality”, it provides rules for “how we can explore it”.  
One then can add axioms to those of QT to get more refined theories, such as Free Quantum Field Theory. The latter can be indeed obtained upon considering a denumerable set of QT systems, and adding the axioms of locality, homogeneity, and isotropy of interactions (see e.g. the review \cite{PWP}).

\newpage
\appendix
\section{Quantum and Classical Theories: axiomatization and main theorems}\label{appendices}
Minimal mathematical axiomatisations of QT and CT as OPTs are provided in Tables \ref{tabminimalQ} and \ref{tabminimalC}. For an OPT we need to provide the mathematical description of  systems, their composition, and transformations from one system to another. Then all rules of compositions of transformations and their respective systems are provided by the OPT framework. The reader who is not familiar with such framework can simply use the intuitive construction of quantum circuits. In Tables
 \ref{tabmainthmsQ}, \ref{tabmainthmsC} we report the main theorems following from the axioms. The reader interested in the motivations for the present  axiomatization is addressed to Ref. \cite{DAriano-nopuri}.


\begin{table}[H]\renewcommand{\arraystretch}{.9}
\begin{tabular}{|l|l|}
\hline
$\Bnd^+(\sH)$  &bounded positive operators over $\sH$\\
$\CP_{\leq}$ &trace-non increasing completely positive map\\
$\CP_{=}$ &trace-preserving completely positive map\\
$\sH$ & Hilbert space over $\Cmplx$\\
$\Cone(\Set)$ &conic hull of $\Set$\\
$\Cone_{\leq 1}(\Set)$ &convex hull of $\{\Set\cup{0}\}$\\
$\Conv(\Set)$ &convex hull of $\Set$\\
$\Eff(\rA)$& set of effects of system $\rA$ \\
$\Eff_1(\rA)$& set of deterministic effects of system $\rA$ \\
$\Mrkv_{\leq}$ &normalization-non-incressing right-stochastic Markov matrices\\
$\Mrkv_1$ &normalization-preserving right-stochastic Markov matrices\\
$\Prm(n)$ &$n\times n$ permutation matrices\\
$(\Reals^n)^+_{\leq1}$&  $\{\vx\in\Reals^{n_\rA}|\vx\geq 0, \vx\leq\bf1\}$: simplex ${\bf S}^{n_\rA+1}$ \\
$(\Reals^n)^+_{1}$& $\{\vx\in\Reals^{n_\rA}|\vx\geq 0, \n{\vx}_1=1\}$: simplex ${\bf S}^{n_\rA}$ \\
$\St(\rA)$ &set of states of system $\rA$\\ 
$\St_1(\rA)$& set of deterministic states of system $\rA$\\ 
$\T(\sH)$ &trace-class operators over $\sH$\\
$\T^+(\sH)$  &trace-class positive operators over $\sH$\\
$\T_{\leq1}^+(\sH)$ &positive sub-unit-trace operators over $\sH$\\
$\T_{=1}^+(\sH)$ &positive unit-trace operators over $\sH$\\
$\Trn(\rA\to\rB)$& set of transformations from system $\rA$ to system $\rB$  \\
$\Trn_1(\rA\to\rB)$& set of deterministic transformations from system $\rA$ to system $\rB$\\
$\Uset(\sH)$ &unitary group over $\sH$\\
\hline
& {\bf Special cases corollaries}\\
\hline
&$\T(\Cmplx)=\Cmplx$,\;\;   $\T^+(\Cmplx)=\Reals^+$,\;\; $\T_{\leq1}^+(\Cmplx)=[0,1]$,\;\; $\T_{=1}^+(\Cmplx)=\{1\}$\\
&$\CP(\T(\sH)\to\T(\Cmplx))=\P(\T(\sH)\to\T(\Cmplx))=\{\Tr[\cdot E],\, E\in\Bnd^+(\sH)\}$
\\
&$\CP(\T(\Cmplx)\to\T(\sH))=\P(\T(\Cmplx)\to\T(\sH))=\T^+(\sH)$
\\
&$\CP_{\leq}(\T(\Cmplx)\to\T(\sH))\equiv\T^+_{\leq 1}(\sH)$\\
&$\CP_{\leq}(\T(\sH)\to\T(\Cmplx))\equiv\{\epsilon(\cdot)=\Tr[\cdot E],\, 0\leq E\leq I\}$\\
&$\Mrkv_{\leq}(n,1)=(\Reals^n)^+_{\leq1}$\\
&$\Mrkv _{1}(n,1)=(\Reals^n)^+_{=1}$\\
\hline
\end{tabular}
\caption{Notation, special-cases corollaries, and common abbreviations.}\label{tnotat}
\end{table}

 \begin{table}[]\renewcommand{\arraystretch}{.9}
\begin{center}
\begin{tabular}{|r| c|l|}
\hline
\multicolumn{3}{c}{\textbf{Quantum Theory}} \\
\hline
system &$\rA$& $\sH_\rA$\\
\hline
system composition &$\rA\rB$& $\sH_{\rA\rB}=\sH_\rA\otimes\sH_\rB$\\
\hline
transformation & $\tT\in\Trn(\rA\to\rB)$ & $\tT\in\CP_{\leq}(\T(\sH_\rA)\to\T(\sH_\rB))$
\\
\hline
Born rule &$p(\tT)=\Tr\tT$& $\tT\in\Trn(\rI\to\rA)$\\
\hline
\end{tabular}
\smallskip
\caption{{\bf Mathematical axiomatisation of Quantum Theory.} As given in the table, in Quantum Theory to each system $\rA$ we associate a Hilbert space over the complex field $\sH_\rA$. To the composition of systems $\rA$ and $\rB$ we associate the tensor product of Hilbert spaces $\sH_{\rA\rB}=\sH_{\rA}\otimes\sH_{\rB}$.  Transformations from system $\rA$ to $\rB$ are described by trace-nonincreasing completely positive (CP) maps from traceclass operators on $\sH_\rA$ to traceclass operators on $\sH_\rB$. Special cases of transformations are those with input trivial system $\rI$ corresponding  to states, whose trace is the preparation probability, the latter providing an efficient Born rule from which one can derive all joint probabilities of any combination of transformations. Everything else is simply special-case corollaries and one realisation theorem: these are reported in Table \ref{tabmainthmsQ}. 
}\label{tabminimalQ}
\end{center}
\end{table}
\begin{table}[H]\renewcommand{\arraystretch}{.9}\footnotesize
\begin{center}
\begin{tabular}{|r| c|l|}
\hline
\multicolumn{3}{c}{\textbf{Quantum theorems}} \\
\hline
trivial system &$\rI$& $\sH_\rI=\Cmplx$\\
\hline
reversible transf. & $\tU=U\cdot U^\dag$ & $U\in\Uset(\sH_\rA)$\\
\hline
determ. transformation & $\tT\in\Trn_1(\rA\to\rB)$ & $\tT\in\CP_{\leq1}(\T(\sH_\rA)\to\T(\sH_\rB))$\\
\hline
parallel composition &  $\tT_1\in\Trn(\rA\to\rB)$, $\tT_2\in\Trn(\rC\to\rD)$ &$\tT_1\otimes\tT_2$ \\
\hline
sequential composition &  $\tT_1\in\Trn(\rA\to\rB)$, $\tT_2\in\Trn(\rB\to\rC)$ &$\tT_2\tT_1$ \\
\hline
\multirow{4}{*}{states}& $\rho\in\St(\rA)\equiv\Trn(\rI\to\rA)$ & $\rho\in\T^+_{\leq1} (\sH_\rA)$\\
\cline{2-3}
  & $\rho\in\St_1(\rA)\equiv\Trn_1(\rI\to\rA)$ & $\rho\in\T^+_{=1} (\sH_\rA)$\\
\cline{2-3}
          & $\rho\in\St(\rI)\equiv\Trn(\rI\to\rI)$ & $\rho\in[0,1]$\\
\cline{2-3}
          & $\rho\in\St_1(\rI)\equiv\Trn(\rI\to\rI)$ & $\rho=1$\\
\hline
\multirow{2}{*}{effects}& $\epsilon\in\Eff(\rA)\equiv\Trn(\rA\to\rI)$ & $\epsilon(\cdot)=\Tr_\rA[\cdot E],\; 0\leq E\leq I_A$\\
\cline{2-3}
  & $\epsilon\in\Eff_1(\rA)\equiv\Trn_1(\rA\to\rI)$ & $\epsilon=\Tr_\rA$\\
\hline
\shortstack{\\Transformations as \\ unitary interaction\\ +\\ von Neumann-Luders}
    &
 $ \!\!\!\!\!\begin{aligned}
      \Qcircuit @C=1em @R=.7em @! R {
        &\poloFantasmaCn{\rA}\qw&\gate{\tT_i}&\poloFantasmaCn{\rB}\qw&\qw}
    \end{aligned} =\!\!\!\!\!\!\!\!\!\!\!\!
    \begin{aligned}
  \Qcircuit
    @C=1em @R=.7em @! R {
&\poloFantasmaCn{\rA}\qw&\multigate{1}{\tU}  &\poloFantasmaCn{\rB}\qw&\qw\\
\prepareC{\sigma}  &\poloFantasmaCn{\rF}\qw&\pureghost{\tU}\qw&\poloFantasmaCn{\rE}\qw&\measureD{\tP_i}
}
\end{aligned}$\!\!
& 
  \shortstack{$\tT_i\rho=\Tr_\rE[U(\rho\otimes\sigma)U^\dag(I_\rB\otimes P_i)]$}
\\
\hline
\end{tabular}
\smallskip
\caption{{\bf Corollaries and a theorem of Quantum Theory, starting from Table \ref{tabminimalQ} axiomatization.}
The first corollary states that the trivial system $\rI$ in order to satisfy the composition rule $\rI\rA=\rA\rI=\rA$ must be associated to the one-dimensional Hilbert space $\sH_\rI=\Cmplx$, since it is the only Hilbert space which trivializes the Hilbert space tensor product. The second corollary states that the reversible transformations are the unitary ones. The third corollary states that the deterministic transformations are the trace-preserving ones. Then the fourth and fifth corollaries give the composition of transformations in terms of compositions of maps. We then have four corollaries about states: 1) states are transformations starting from the trivial system and, as such, are positive operators on the system Hilbert space, having trace bounded by one; 2) the deterministic states correspond to unit-trace positive operator;  3) the states of trivial system are just probabilities; 4) The only trivial system deterministic state is the number 1. We then have two corollaries for effects, as special cases of transformation toward the trivial system: 1) the effect is represented by the partial trace over the system Hilbert space of the multiplication with a positive operator bounded by the identity over the system Hilbert space; 2) the only deterministic effect is the partial trace over the system Hilbert space. Finally, we have the realization theorem for transformations in terms of unitary interaction $\tU=U\cdot U^\dag$ with an environment $\rF$ and a projective effect-test $\{\tP_i\}$ over environment $\rE$, with $\tP_i=P_i\cdot P_i$, $\{P_i\}$ being a complete set of orthogonal projectors.
}\label{tabmainthmsQ}
\end{center}
\end{table}

\begin{table}[H]\renewcommand{\arraystretch}{.9}
\begin{center}
\begin{tabular}{|r| c|l|}
\hline
\multicolumn{3}{c}{\textbf{Classical Theory}} \\
\hline
system &$\rA$& $\Reals^{n_\rA}$\\
\hline
system composition &$\rA\rB$& $\Reals^{n_{\rA\rB}}=\Reals^{n_\rA}\otimes \Reals^{n_\rB}$\\
\hline
transformation & $\tT\in\Trn(\rA\to\rB)$ & $\tT\in\Mrkv_{\leq}(\Reals^{n_\rA}, \Reals^{n_\rB})$\\
\hline
\end{tabular}\smallskip
\caption{{\bf Mathematical axiomatisation of Classical Theory.} To each system $\rA$ we associate a real Euclidean space $\Reals^{n_\rA}$. To composition of systems $\rA$ and $\rB$ we associate the tensor product spaces $\Reals^{n_\rA}\otimes \Reals^{n_\rB}$.  Transformations from system $\rA$ to system $\rB$ are described by substochastic Markov matrices from the input  space to the output  space. Everything else are simple special-case corollaries: these are reported in Table \ref{tabmainthmsC}. 
}\label{tabminimalC}
\end{center}
\end{table}

\begin{table}[H]\renewcommand{\arraystretch}{.9}\footnotesize
\begin{center}
\begin{tabular}{|r| c|l|}
\hline
\multicolumn{3}{c}{\textbf{Classical theorems}} \\
\hline
trivial system &$\rI$& $\Reals$\\
\hline
reversible transformations & $\tP$  & $\tP\in\Prm(n_\rA)$\\
\hline
transformation & $\tT\in\Trn_\leqslant (\rA\to\rB)$ & $
\tT\in\Mrkv_\leqslant(\Reals^{n_\rA},\Reals^{n_\rB})$\\
\hline
determ. transformation & $\tT\in\Trn_1(\rA\to\rB)$ & $
\tT\in\Mrkv_1(\Reals^{n_\rA},\Reals^{n_\rB})$\\
\hline
parallel composition &  $\tT_1\in\Trn(\rA\to\rB)$, $\tT_2\in\Trn(\rC\to\rD)$ &$\tT_1\otimes\tT_2$ \\
\hline
sequential composition &  $\tT_1\in\Trn(\rA\to\rB)$, $\tT_2\in\Trn(\rB\to\rC)$ &$\tT_2\tT_1$ \\
\hline
\multirow{4}{*}{states}& $\vx\in\St(\rA)\equiv\Trn(\rI\to\rA)$ & $\vx\in(\Reals^{n_\rA})^+_{\leq1}$\\
\cline{2-3}
& $\vx\in\St_1(\rA)\equiv\Trn_1(\rI\to\rA)$ & $\vx\in(\Reals^{n_\rA})^+_{=1}$\\
\cline{2-3}
          & $p\in\St(\rI)\equiv\Trn(\rI\to\rI)$ & $p\in[0,1]$\\
\cline{2-3}
          & $p\in\St_1(\rI)\equiv\Trn(\rI\to\rI)$ & $p=1$\\
\hline
\multirow{2}{*}{effects}& $\epsilon\in\Eff(\rA)\equiv\Trn(\rA\to\rI)$ & $\epsilon(\cdot)=\cdot \vx,\; {\bf 0}\leq \vx\leq {\bf 1}$\\
\cline{2-3}
  & $\epsilon\in\Eff_1(\rA)\equiv\Trn_1(\rA\to\rI)$ & $\epsilon=\cdot {\bf 1}$\\
\hline
\end{tabular}
\caption{{\bf Main theorems of Classical Theory, starting from axioms in Table \ref{tabminimalC}}.
The first corollary states that the trivial system $\rI$ in order to satisfy the composition rule $\rI\rA=\rA\rI=\rA$ must be associated to the one-dimensional space $\Reals$, since it is the only real linear space that trivialises the tensor product. The second corollary states that the reversible transformations are the permutation matrices. The third states that transformations are substochastic Markov matrices. The fourth states that the deterministic transformations are stochastic Markov matrices. Then the fifth and sixth corollaries give the composition of transformations in terms of composition of matrices. We then have four corollaries about states: 1) states are transformations starting from the trivial system and, as such, are sub-normalized probability vectors (vectors in the positive octant with sum of elements bounded by one; 2) the deterministic states correspond to normalised probability vectors;  3) the case of trivial output-system correspond to just probabilities; 4) The only trivial output-system deterministic state is the number 1. We then have two corollaries for effects, as special cases of transformation toward the trivial system: 1) the effect is represented by scalar product with a vector with components in the unit interval; 2) the only deterministic effect is the scalar product with the vector with all unit components.
}\label{tabmainthmsC}
\end{center}
\end{table}

\begin{acknowledgements} I thank Arkady Plotnitsky for enjoyable interesting discussions about quantum theory.
\end{acknowledgements}

\end{document}